\documentclass[12pt]{article}
\usepackage{epsfig,amsmath,amsfonts,amssymb,amstext,afterpage,psfrag}
\long\def\symbolfootnote[#1]#2{\begingroup%
\def\thefootnote{\fnsymbol{footnote}}\footnote[#1]{#2}\endgroup}

\def\spose#1{\hbox to 0pt{#1\hss}}
\def\lsim{\mathrel{\spose{\lower 3pt\hbox{$\mathchar"218$}}
 \raise 2.0pt\hbox{$\mathchar"13C$}}}
\def\gsim{\mathrel{\spose{\lower 3pt\hbox{$\mathchar"218$}}
 \raise 2.0pt\hbox{$\mathchar"13E$}}}

\catcode`@=11
\def\@citex[#1]#2{%
  \if@filesw\immediate\write\@auxout{\string\citation{#2}}\fi
  \def\@citea{}\@cite{\@for\@citeb:=#2\do
    {\@citea\def\@citea{,\penalty\@m}\@ifundefined
      {b@\@citeb}{{\bf ?}\@warning
{Citation `\@citeb' on page \thepage \space undefined}}%
      \hbox{\csname b@\@citeb\endcsname}}}{#1}}
\def\citer{\@ifnextchar [{\@tempswatrue\@citexr}{\@tempswafalse\@citexr[]}}
  \def\@citexr[#1]#2{%
    \if@filesw\immediate\write\@auxout{\string\citation{#2}}\fi
    \def\@citea{}\@cite{\@for\@citeb:=#2\do
      {\@citea\def\@citea{--\penalty\@m}\@ifundefined
{b@\@citeb}{{\bf ?}\@warning
{Citation `\@citeb' on page \thepage \space undefined}}%
\hbox{\csname b@\@citeb\endcsname}}}{#1}}

\begin{document}

\begin{titlepage}

\begin{flushright}
{\small
LMU-ASC~70/13\\ 
}
\end{flushright}

\vspace{0.5cm}
\begin{center}
{\Large\bf \boldmath                                               
Nonstandard Higgs Couplings\\
\vspace*{0.3cm}                                                            
from Angular Distributions in $h\to Z\ell^+\ell^-$ 
\unboldmath}
\end{center}

\vspace{0.5cm}
\begin{center}
{\sc Gerhard Buchalla$^1$, Oscar Cat\`a$^{1,2,3}$ and Giancarlo D'Ambrosio$^4$} 
\end{center}

\vspace*{0.4cm}

\begin{center}
$^1$Ludwig-Maximilians-Universit\"at M\"unchen, Fakult\"at f\"ur Physik,\\
Arnold Sommerfeld Center for Theoretical Physics, 
80333 M\"unchen, Germany\\
\vspace*{0.1cm}
$^2$TUM-IAS, Lichtenbergstr. 2a, D-85748 Garching, Germany\\
\vspace*{0.1cm}
$^3$Physik Department, TUM, D-85748 Garching, Germany\\
\vspace*{0.1cm}
$^4$INFN-Sezione di Napoli, Via Cintia, 80126 Napoli, Italy
\end{center}

\vspace{1.5cm}
\begin{abstract}
\vspace{0.2cm}\noindent
We compute the fully differential rate for the Higgs-boson decay
$h\to Z\ell^+\ell^-$, with $Z\to\ell^{'+}\ell^{'-}$.  For these processes
we assume the most general matrix elements within an
effective Lagrangian framework. The electroweak chiral Lagrangian
we employ assumes minimal particle content and Standard Model gauge
symmetries, but is otherwise completely general. We discuss how information
on new physics in the decay form factors may be obtained that is 
inaccessible in the dilepton-mass spectrum integrated over angular
variables. The form factors are related to the coefficients of the
effective Lagrangian, which are used to estimate the potential size of
new-physics effects.
\end{abstract}

\vfill

\end{titlepage}

\section{Introduction}
\label{sec:intro}

The recent discovery of a light scalar $h$ by ATLAS~\cite{Aad:2012tfa} and 
CMS~\cite{Chatrchyan:2012ufa} has been a major step forward in our 
understanding of electroweak symmetry breaking. The first run of the LHC has 
established its mass with an accuracy of better than $1\%$ and has provided 
evidence for its scalar nature with spin-parity $0^+$~\cite{Aad:2013xqa}. 
Furthermore, decay rates to gauge boson pairs show no significant deviations 
from their Standard Model (SM) values~\cite{Djouadi:2005gi,Heinemeyer:2013tqa} 
within the present accuracy of around 
$20-30\%$~\cite{ATLAS:2013sla,CMS:yva}. The overall agreement with the 
Standard Model is so far impressive.

However, theoretical arguments suggest that deviations should be expected. 
Their absence would actually be rather puzzling and would point to a 
fine-tuned solution for electroweak symmetry breaking, where the lightness of 
the Higgs would remain unexplained. 
Deviations from the Standard Model parameters open the gate to 
new physics, expected to lie at the Terascale in the form of weakly or 
strongly-coupled new interactions. So far the LHC has been able to test total 
decay rates of $h$ into gauge boson pairs. However, LHC run~2, with a 
substantial increase in luminosity, will provide enough statistics to probe 
also differential distributions, thereby testing the Standard Model in much 
greater detail.

In this paper we will study in a model-independent way the impact of new 
physics in the full angular distribution of $h\to Z{\ell}^+{\ell}^-$ decay, 
with the $Z$ on-shell and eventually decaying into a lepton pair. 
We will argue that $h\to Z{\ell}^+{\ell}^-$ is a useful channel not only 
for spin identification~\citer{Choi:2002jk,Modak:2013sb}, 
but also to test nonstandard couplings: it provides a rich 4-body angular 
distribution with a clean 4-lepton final-state signature. For earlier work 
see~\cite{Barger:1993wt,Stolarski:2012ps}. 

Our results can be parametrized in terms of 6 independent dynamical form 
factors, which include the effects of virtual electroweak bosons 
($\gamma$ and $Z$) as well as heavier states, whose effects at the electroweak 
scale are encoded in contact interactions. Since we aim at model independence, 
we will study the new physics contributions using the effective field theory
(EFT) scheme developed in~\cite{Buchalla:2012qq,Buchalla:2013rka}, which is the most general 
EFT of the electroweak interactions. 
As opposed to particular models, the resulting set of new 
physics coefficients will remain undetermined. However, their natural sizes 
can still be estimated with the aid of power-counting arguments.

Certain aspects of this decay mode have already been discussed 
recently~\citer{Isidori:2013cla,Isidori:2013cga}, with a focus on the
dilepton-mass distribution. The observation there is that mass distributions 
can unveil 
new physics structures in an otherwise SM-compatible integrated decay rate. 
This however comes at the expense of some fine-tuning in the new physics 
parameters. In contrast, by exploiting angular distributions one can 
identify structures that do not contribute to the integrated decay 
rate. Thus, one can still be compatible with the SM decay rates without 
tuning the new-physics parameters.

As opposed to loop-induced processes, such as $h\to\gamma Z$, 
$h\to Z\ell^+\ell^-$ does not look a priori like a promising testing ground 
for new-physics effects. As we will show below, they are 
expected, at most, at the few $\%$ level in certain observables. 
$h\to Z\ell^+\ell^-$ is however
an exceptionally clean decay mode and the natural suppression of new physics can be compensated with 
statistics. In fact, the LHC running at 14 TeV with an integrated luminosity of 3000 fb$^{-1}$ will potentially
be sensitive to new-physics effects in $h\to Z\ell^+\ell^-$. Our analysis also shows that CP-odd
effects in $h\to Z\ell^+\ell^-$ are expected only at the per-mille level.

The remainder of this paper will be organized as follows: in Section 2 we 
will derive the full angular distribution for $h\to Z{\ell}^+{\ell}^-$. 
Expressions for the dynamical form factors in terms of EFT coefficients will 
be given in Section 3, with a discussion of their expected sizes in both 
weakly and strongly-coupled scenarios. In Section 4 we will discuss some 
selected angular observables. Conclusions 
are given in Section 5 while an appendix with kinematical details is provided 
for reference.


\section{Angular distribution for $h\to Z{\ell}^+{\ell}^-$}
\label{sec:basfor}

We denote the amplitude for the $h\to Z \ell^+\ell^-$ decay of a Higgs boson
by $\varepsilon^\mu {\cal M}_{3,\mu}$, and the decay of an on-shell $Z$-boson
into a lepton pair by $\varepsilon^\mu {\cal M}_{2,\mu}$, 
where $\varepsilon^\mu$ is the $Z$-boson polarization.
The fully differential decay rate for $h(k)\to Z(p)\ell^+(q_1)\ell^-(q_2)$, 
followed by $Z(p)\to \ell^{'+}(p_1)\ell^{'-}(p_2)$, is then given, in the narrow-width approximation, by
\begin{equation}\label{dgamma}
\frac{d\Gamma}{ds\, d\cos\alpha\, d\cos\beta\, d\phi} =
\frac{\lambda}{(2\pi)^5\, 2^{10}\, \sqrt{r}\Gamma_Z}
\left| {\cal M}^\mu_3 {\cal M}_{2,\mu}\right|^2
\end{equation}
where we have defined
\begin{equation}\label{rslamdef}
r=\frac{m^2_Z}{M^2_h}\, ,\quad s=\frac{q^2}{M^2_h}\, ,\quad
\lambda=(1+ r^2 + s^2 - 2r - 2s - 2 r s)^{1/2}
\end{equation}
and $\Gamma_Z$ is the total width of the $Z$. The kinematics is further discussed in Appendix \ref{sec:app}. 

For massless leptons the decay amplitudes can be written as 
($\epsilon_{0123}=+1$)
\begin{align}\label{m3def}
&{\cal M}_{3,\mu} = i \frac{2^{1/4} G^{1/2}_F r}{s-r} \\
& \quad\cdot\bar u(q_2)\left[
2 F_1 \gamma_\mu(G_V - G_A \gamma_5) + 
\frac{q_\mu}{M^2_h} \not\! k (H_V - H_A \gamma_5) +
\frac{\epsilon_{\alpha\mu\beta\lambda}}{M^2_h} p^\alpha q^\beta
\gamma^\lambda(K_V - K_A \gamma_5)\right] v(q_1)\nonumber
\end{align}
and
\begin{equation}\label{m2def}
{\cal M}_{2,\mu} = i \bar u(p_2)\gamma_\mu(g_V - g_A \gamma_5) v(p_1)
\end{equation}
The form of the amplitude in (\ref{m3def}) is valid through 
next-to-leading order (NLO) of the general electroweak effective
Lagrangian described in \cite{Buchalla:2013rka} and in Section \ref{sec:leff}.
The form factors $G_{V,A}$, $H_{V,A}$, $K_{V,A}$ are functions of 
$r$ and $s$. The global normalization of the amplitude has been chosen such
that in the Standard Model at leading order $F_1\equiv 1$,
$G_V=g_V$, $G_A=g_A$.

Summing over the final-state lepton polarizations gives
\begin{equation}\label{m3m2jj}
\left| {\cal M}^\mu_3 {\cal M}_{2,\mu}\right|^2 = \sqrt{2} G_F M^4_h
\left(\frac{r}{r-s}\right)^2\, J(r,s,\alpha,\beta,\phi)
\end{equation}
where
\begin{eqnarray}\label{jdef}
J(r,s,\alpha,\beta,\phi) &=& J_1\frac{9}{40}(1+\cos^2\alpha \cos^2\beta)+ 
J_2\frac{9}{16}\sin^2\alpha \sin^2\beta + J_3 \cos\alpha \cos\beta
\nonumber\\
&& +\left( J_4 \sin\alpha \sin\beta + 
       J_5 \sin 2\alpha \sin 2\beta\right) \sin\phi \nonumber\\
&& +\left( J_6 \sin\alpha \sin\beta + 
       J_7 \sin 2\alpha \sin 2\beta\right) \cos\phi \nonumber\\
&& + J_8 \sin^2\alpha \sin^2\beta \sin 2\phi 
   + J_9 \sin^2\alpha \sin^2\beta \cos 2\phi
\end{eqnarray}
The previous expression factors out the angular dependence, $J_i$ being 
dynamical functions which depend only on the invariant masses $r$, $s$. They are given by
\begin{align}\label{jji}
J_1 &=\frac{640}{9} F^2_1 (G^2_V + G^2_A) (g^2_V + g^2_A) r s\nonumber\\
J_2 &=\frac{32}{9} F_1 (g^2_V + g^2_A) 
  \left[2 F_1(G^2_V + G^2_A) (\lambda^2 + 2 r s)
   +(G_V H_V + G_A H_A)\lambda^2 (1-r-s)\right]\nonumber\\
J_3 &=128 F^2_1 G_V G_A g_V g_A r s\nonumber\\
J_4 &=8 F_1 (G_V K_A + G_A K_V)g_V g_A \lambda
 \sqrt{r s} (1-r-s)\nonumber\\
J_5 &=F_1 (G_V K_V + G_A K_A) (g^2_V + g^2_A) \lambda
 \sqrt{r s} (1-r-s)\nonumber\\
J_6 &=-8F_1 g_V g_A \sqrt{r s} 
  \left[8 F_1 G_V G_A (1-r-s)+(G_V H_A + G_A H_V)\lambda^2\right]\nonumber\\
J_7 &=-F_1 (g^2_V + g^2_A) \sqrt{r s} \left[4 F_1(G^2_V + G^2_A) (1-r-s)+
   (G_V H_V + G_A H_A)\lambda^2\right]\nonumber\\
J_8 &=-4 F_1 (G_V K_V + G_A K_A) (g^2_V + g^2_A) \lambda r s\nonumber\\
J_9 &=8 F^2_1 (G^2_V + G^2_A) (g^2_V + g^2_A) r s
\end{align}

As will be explained in more detail in Section \ref{sec:leff},
the form factors
$G_{V,A}$ receive leading-order contributions in the Standard Model, whereas 
$H_{V,A}$ and $K_{V,A}$ only arise as next-to-leading order corrections and 
capture, respectively, CP-even and CP-odd contributions.
In writing the expression for the $J_i$, we have therefore consistently
neglected terms of second order in $H_{V,A}$ and $K_{V,A}$.
It follows that to leading order in the Standard Model the observables
$J_4$, $J_5$ and $J_8$, which carry the dependence on $K_{V,A}$, are zero, as one
would expect from general CP considerations.

With sufficient data, a general fit to the angular distribution
of the four final-state leptons could in principle extract
all 9 terms $J_i$ in the fully differential decay rate 
(\ref{dgamma}), (\ref{m3m2jj}) and (\ref{jdef}).
From (\ref{jji}) we see that measuring $J_1$, $\ldots$, $J_6$,
for example, would determine the 6 independent combinations
\begin{align}
&G^2_V+ G^2_A,&\qquad &G_V G_A\nonumber\\
&G_V H_V + G_A H_A,&\qquad &G_V H_A + G_A H_V\nonumber\\
&G_V K_V + G_A K_A,&\qquad &G_V K_A + G_A K_V
\end{align}  
All of the 6 form factors $G_{V,A}$, $H_{V,A}$, $K_{V,A}$ could then
be obtained. The remaining three observables $J_7$, $J_8$, $J_9$
give no independent information on these form factors.
They can be used for a cross-check or as alternative input.
The coefficients $g_{V,A}$ are not directly related to the process
$h\to Z\ell^+\ell^-$ and have to be constrained independently from properties 
of $Z$ decays.

With limited data, it is more efficient to extract the different
$J_i$ projecting them from (\ref{jdef}). Integrating the distribution 
in (\ref{jdef}) over $\phi$ we are left with $J_1$, $J_2$ and $J_3$ as 
the only observables. Integrating in addition
over $\alpha$ and $\beta$ eliminates $J_3$. Thus, the differential rate
$d\Gamma/ds$, fully integrated over the angular variables, 
remains sensitive only to $J_1+J_2$. Performing the angular integrations one 
obtains the dilepton-mass spectrum of the $h\to Z\ell^+\ell^-$ rate, multiplied 
by the $Z\to \ell^{'+}\ell^{'-}$ branching fraction $B_\ell$. From (\ref{dgamma}) 
one finds
\begin{align}\label{dgamds}
\frac{d\Gamma}{ds} &=B_\ell\, \frac{G_F M^3_h}{\sqrt{2}\, 192\pi^3}
\frac{\lambda r}{(r-s)^2}\\ 
&\quad \times F_1 \left[2 F_1(G^2_V + G^2_A) (\lambda^2 + 12 r s)
   +(G_V H_V + G_A H_A)\lambda^2 (1-r-s)\right]\nonumber
\end{align}
where
\begin{equation}\label{bl}
B_\ell=\frac{(g^2_V + g^2_A) m_Z}{12\pi \Gamma_Z}
\end{equation} 

In contrast, $J_3$, $\ldots$, $J_9$ have to be accessed with appropriate 
angular asymmetries. For instance, the term $J_3$ can be extracted
by integrating over $\phi$ and forming a suitable forward-backward 
asymmetry in $\cos\alpha$ and $\cos\beta$. In Section~\ref{sec:III} we 
examine this and other angular asymmetries in detail.

The angular distribution in $h\to Z\ell^+\ell^-$ is similar to
the one in the rare $B$-meson decay $B\to K^*\ell^+\ell^-$, which has been 
discussed for instance in \citer{Egede:2008uy,Bobeth:2008ij}. However,
in the present case the angles $\alpha$ and $\beta$ are on an equal footing, 
and accordingly the angular dependence in (\ref{jdef}) 
is symmetric under the interchange of $\alpha$ and $\beta$.
Note in particular that the forward-backward asymmetry term $J_3$
is proportional to the product $\cos\alpha\, \cos\beta$, thus
representing a kind of correlated double asymmetry in $\alpha$ and $\beta$. 
It vanishes when either $\alpha$ or $\beta$ are integrated over
their full range. This is in contrast to $B\to K^*\ell^+\ell^-$,
where a forward-backward asymmetry in the single angle $\alpha$
exists due to the more complicated structure of the 
hadronic transition $B\to K^*$.


\section{Form factors from effective Lagrangian}
\label{sec:leff}

In order to estimate the form factors $G_{V,A}$, $H_{V,A}$, $K_{V,A}$ 
in (\ref{m3def}) and $g_{V,A}$ in (\ref{m2def}) we will work with the 
nonlinear effective Lagrangian discussed in \cite{Buchalla:2012qq,Buchalla:2013rka}. A subset
of the relevant operators has also been discussed in~\cite{Alonso:2012px,Alonso:2012pz}. 
In this framework, electroweak symmetry breaking is realized by spontaneously 
breaking a global $SU(2)_L\times SU(2)_R$ down to $SU(2)_V$. The resulting 
Goldstone modes are then collected into a matrix $U$ transforming as 
$g_L U g_R^{\dagger}$ under the global group. One also defines
\begin{align}
D_\mu U=\partial_\mu U+i g W_\mu U -i g' B_\mu U T_3
\end{align}
such that the SM subgroup $SU(2)_L\times U(1)_Y$ is gauged. 
For convenience we will use the shorthand notation \begin{align}
L_{\mu}=iUD_{\mu}U^{\dagger},\qquad \tau_L=UT_3U^{\dagger}
\end{align}
for the Goldstone covariant derivative and the custodial symmetry breaking 
spurion $T_3$. 
The Higgs field $h$ is introduced as an additional light (pseudo-Goldstone)
boson, singlet under the SM gauge group.

With these definitions one has at leading 
order~\cite{Buchalla:2013rka,Contino:2010mh,Contino:2010rs}
\begin{align}\label{LO}
{\cal{L}}_{LO}=&-\frac{1}{2}\langle W_{\mu\nu}W^{\mu\nu}\rangle 
-\frac{1}{4} B_{\mu\nu}B^{\mu\nu} + i\!\sum_{f}\bar {\psi}_f \!\not\!\! D\psi_f
\nonumber\\
&+\frac{v^2}{4}\ \langle L_\mu L^\mu\rangle f\left(\frac{h}{v}\right)
-\frac{1}{2}h(\partial^2+M^2_h)h - V(h)
\end{align}   
For $h\to Z{\ell}^+{\ell}^-$ the final-state fermions can be taken massless
to an excellent approximation and therefore we have omitted the Yukawa 
terms above. The main contribution to $h\to Z\ell^+\ell^-$ comes from the 
subprocess $h\to ZZ^*$, which is described by the gauge-boson mass term, where 
$f(h/v)$ can be truncated at linear order for the process of interest here:
\begin{align}
f\left(\frac{h}{v}\right)=1+2a\frac{h}{v}
\end{align}
At next-to-leading order (NLO) there are 8 relevant CP-even operators
\begin{align}\label{naive1}
{\cal{O}}_{Xh1}&=g^{\prime 2} B_{\mu\nu}B^{\mu\nu}\frac{h}{v} 
f_{Xh1}\left(\frac{h}{v}\right),&\qquad
&{\cal{O}}_{Xh2}=g^2\langle W_{\mu\nu}W^{\mu\nu}\rangle\frac{h}{v} 
f_{Xh2}\left(\frac{h}{v}\right)\nonumber\\
{\cal{O}}_{XU1}&=g^{\prime}gB_{\mu\nu}\langle W^{\mu\nu}\tau_L\rangle 
f_{XU1}\left(\frac{h}{v}\right),&\qquad 
&{\cal{O}}_{XU2}=g^2 \langle W^{\mu\nu}\tau_L\rangle^2 
f_{XU2}\left(\frac{h}{v}\right)\nonumber\\
{\cal{O}}_{V7}&=-{\bar{l}}\gamma_{\mu}l\langle \tau_LL^{\mu}\rangle 
f_{V7}\left(\frac{h}{v}\right),&\qquad
&{\cal{O}}_{V8}=-{\bar{l}}\gamma_{\mu}\tau_L l\langle \tau_LL^{\mu}\rangle 
f_{V8}\left(\frac{h}{v}\right)\nonumber\\
{\cal{O}}_{V10}&=-{\bar{e}}\gamma_{\mu}e\langle \tau_LL^{\mu}\rangle 
f_{V10}\left(\frac{h}{v}\right),&\qquad &{\cal{O}}_{\beta_1}=
-v^2\langle\tau_LL_{\mu}\rangle^2f_{\beta_1}\left(\frac{h}{v}\right)
\end{align}
and 4 CP-odd ones:
\begin{align}\label{naive2}
{\cal{O}}_{Xh4}&=g^{\prime 2} \epsilon_{\mu\nu\lambda\rho}B^{\mu\nu} B^{\lambda\rho} 
\frac{h}{v} f_{Xh3}\left(\frac{h}{v}\right),&\quad
&{\cal{O}}_{Xh5}=g^2\epsilon_{\mu\nu\lambda\rho}\langle W^{\mu\nu} W^{\lambda\rho}\rangle 
\frac{h}{v} f_{Xh4}\left(\frac{h}{v}\right)\nonumber\\
{\cal{O}}_{XU4}&=g^{\prime}g \epsilon_{\mu\nu\lambda\rho} B^{\mu\nu}
\langle W^{\lambda\rho}\tau_L\rangle f_{XU4}\left(\frac{h}{v}\right),&\quad
&{\cal{O}}_{XU5}=g^2\epsilon_{\mu\nu\lambda\rho} \langle\tau_L W^{\mu\nu}\rangle
\langle\tau_L W^{\lambda\rho}\rangle f_{XU5}\left(\frac{h}{v}\right)
\end{align}
Some comments are in order:
\begin{itemize}
\item Fermionic tensor operators are in principle also present, but they turn out 
to be negligible: first, they have a chiral suppression and second, they 
do not interfere with the Standard Model and thus can only appear at NNLO.
\item For simplicity, the list above includes only fermions of the first family. The extension
to include the second family is however trivial.
\item The $f_i(h/v)$ above are generic functions with model-dependent 
coefficients~\cite{Buchalla:2013rka}. As a result, the previous operators contain all the 
possible powers of $h$. In the following, $a_i$ and $b_i$ will denote, 
respectively, the dimensionless Wilson coefficients for the pieces without $h/v$ 
and linear in $h/v$, which are the relevant ones for the process under study.
\end{itemize}
\begin{figure}[t]
\begin{center}
\includegraphics[width=5.1cm]{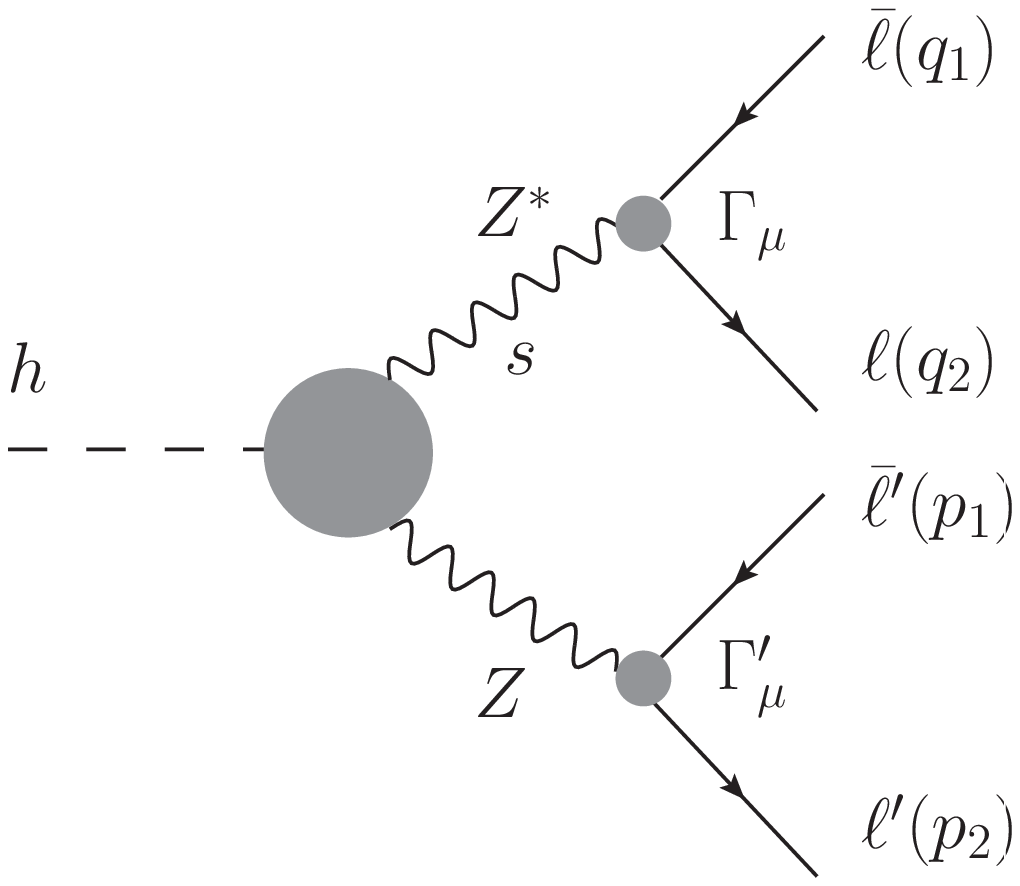}
\includegraphics[width=5.1cm]{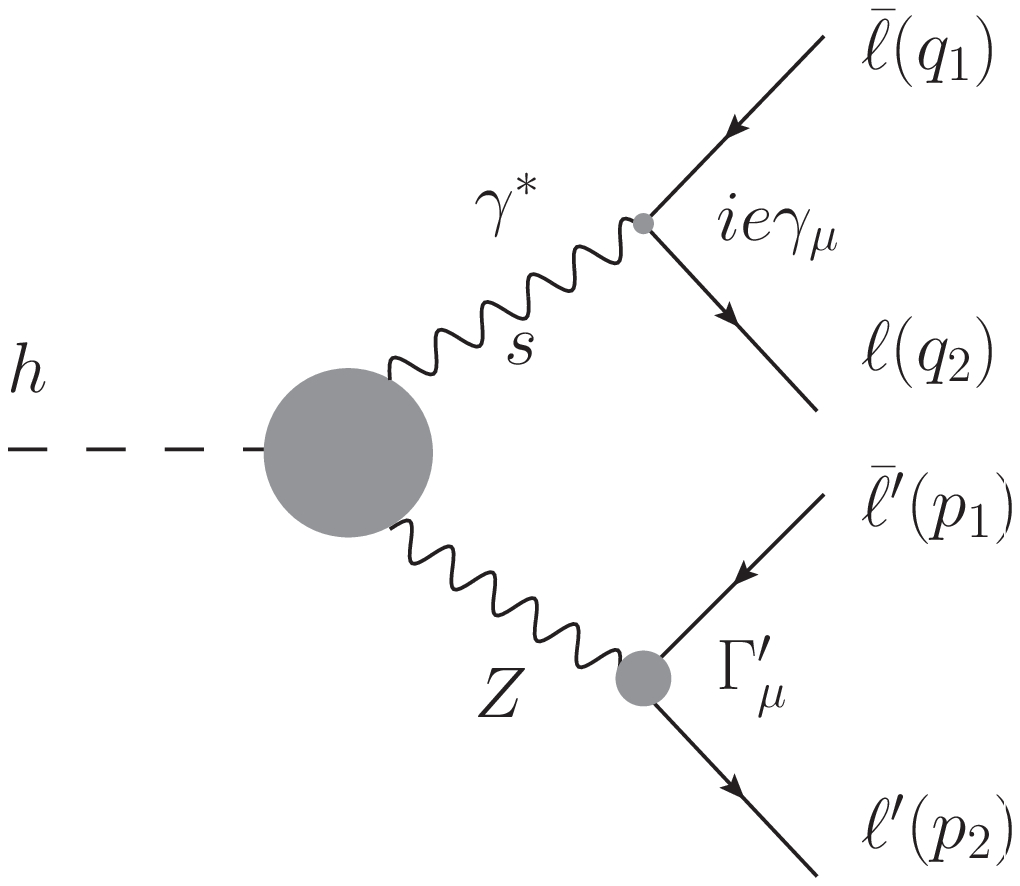}
\includegraphics[width=5.1cm]{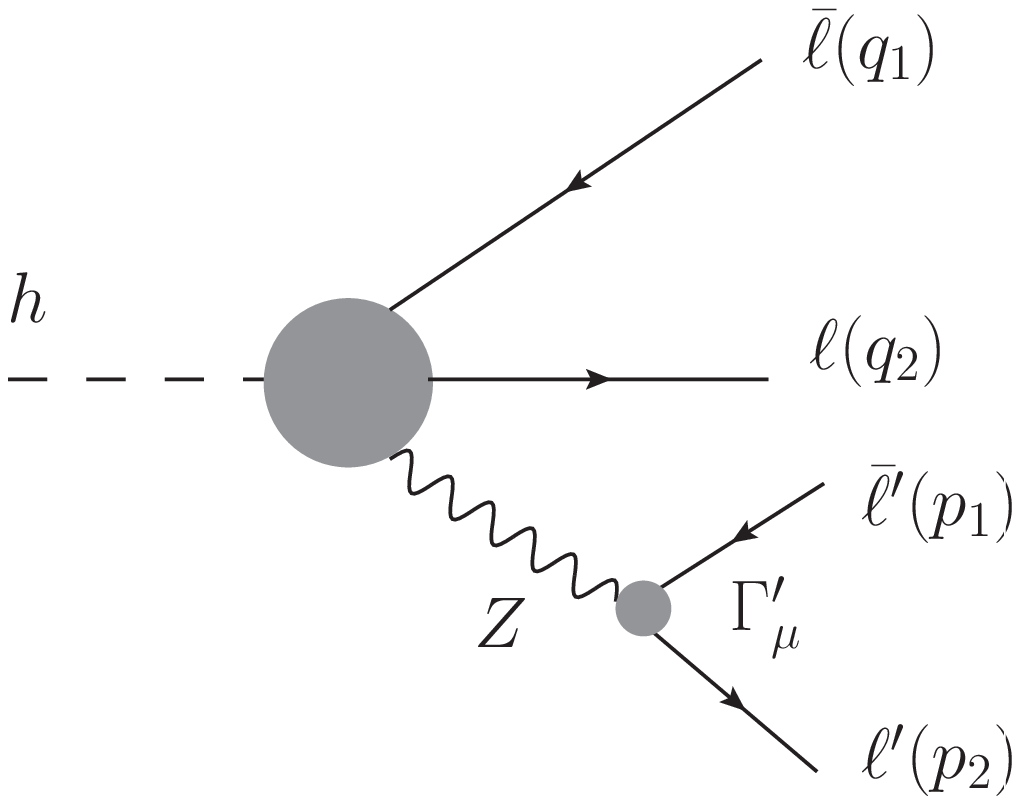}
\end{center}
\caption{\small{\it{Different contributions to $h\to Z\ell^+\ell^-$.}}}
\label{fig:1}
\end{figure}

The operators above give the most general direct contributions to the $h\to ZZ$ 
and $Z\to \ell^+\ell^-$ vertices, but also lead to a renormalization of the
fields and parameters~\cite{Holdom:1990xq,Buchalla:2013wpa}. These effects will be
consistently included in all our results. As the fundamental electroweak parameters we will employ 
$\alpha_{em}=e^2/4\pi$, $m_Z$ and $G_F$ ($Z$-standard definition). Then the NLO 
corrections can finally be expressed in terms of the following effective 
interactions
\begin{align}\label{effective}
{\cal{L}}_{NLO} = & 2^{1/4} G^{1/2}_F m_Z^2F_1\, h\, Z^{\mu}Z_{\mu}\nonumber\\
&+b_2 \frac{h}{v}Z^{\mu\nu}Z_{\mu\nu}+b_2^{\gamma}\frac{h}{v}Z^{\mu\nu}A_{\mu\nu}
+b_3 \frac{h}{v} \epsilon_{\mu\nu\lambda\rho} Z^{\mu\nu} Z^{\lambda\rho}+
b_3^{\gamma}\frac{h}{v} \epsilon_{\mu\nu\lambda\rho} Z^{\mu\nu}A^{\lambda\rho}\nonumber\\
&+Z_\mu{\bar{l}}\gamma^{\mu}\bigg[g_V - g_A\gamma_5\bigg]l+
\frac{h}{v} Z_\mu{\bar{l}}\gamma^{\mu}\bigg[h_V - h_A\gamma_5\bigg]l
\end{align}
For convenience, we have defined
\begin{align}\label{gvahva}
g_V &=\frac{g}{4 c_Z}(\kappa_1 - 4 s^2_Z\kappa_2),
\qquad g_A=\frac{g}{4 c_Z}\kappa_1
\end{align}
such that at leading order in the Standard Model $\kappa_i=1$. 
Here $s_Z$ ($c_Z$) denotes the sine (cosine) of the Weinberg angle in the
$Z$-standard definition ($\alpha=\alpha(m_Z)$)
\begin{equation}\label{szcz}
s^2_Z c^2_Z\equiv\frac{\pi\alpha}{\sqrt{2} G_F m^2_Z}
\end{equation}
and $g$ is the $SU(2)_L$ gauge coupling, where $g s_Z = e=\sqrt{4\pi\alpha}$.
By analogy to (\ref{gvahva}) we have defined
\begin{align}
h_V &=\frac{g}{4 c_Z}(\omega_1 - 4 s^2_Z\omega_2),
\qquad h_A=\frac{g}{4 c_Z}\omega_1
\end{align}
In terms of the coefficients of (\ref{effective}), the form factors read
\begin{align}\label{gvaleff}
G_V &= g_V \left(1-\frac{b_2}{F_1}\frac{1-r-s}{r}\right)
-\frac{b_2^{\gamma} e}{2 F_1}\frac{(1-r-s)(s-r)}{r s}  + 
\frac{h_V}{2 F_1} \frac{s-r}{r}
\nonumber\\
G_A &= g_A \left(1-\frac{b_2}{F_1}\frac{1-r-s}{r}\right)
+ \frac{h_A}{2 F_1} \frac{s-r}{r}\nonumber\\
H_V &= \frac{4 b_2}{r} g_V + 2 b_2^{\gamma} e\frac{s-r}{r s}\, ,\qquad
H_A = \frac{4 b_2}{r} g_A\nonumber\\
K_V &= -\frac{8 b_3}{r} g_V - 4 b_3^{\gamma} e\frac{s-r}{r s}\, ,\qquad
K_A = -\frac{8 b_3}{r} g_A
\end{align}
In turn, the operators in the Lagrangian (\ref{effective}) can be expressed in terms 
of the basic EFT operators listed before. 
For the coefficients this implies ($t_Z=s_Z/c_Z$), 
\begin{align}\label{corrections1}
F_1&=a(1+ 2\beta_1 -\delta_G) - b_{\beta_1}\nonumber\\
b_{2,3}&=\frac{e^2}{2}\left(2t_Z^2b_{Xh1,4}+t_Z^{-2}b_{Xh2,5}-b_{XU1,4}+
\frac{t_Z^{-2}}{2}b_{XU2,5}\right)\nonumber\\
b_{2,3}^{\gamma}&=e^2\left(-2t_Z b_{Xh1,4}+t_Z^{-1}b_{Xh2,5}+
\frac{1}{2}(t_Z^{-1} -t_Z) b_{XU1,4}+\frac{t_Z^{-1}}{2}b_{XU2,5}\right)
\end{align}
and
\begin{align}\label{omdef1}
\kappa_1&\equiv 1-a_{V7}+\frac{1}{2} a_{V8} + a_{V10}+ \beta_1-\delta_G \nonumber\\
\kappa_2&\equiv 1+\frac{1}{2s^2_Z} a_{V10}
+\frac{\delta_G-\beta_1 - a_{XU1} e^2/s^2_Z}{c^2_Z - s^2_Z}
\nonumber\\
\omega_1&\equiv -b_{V7}+\frac{1}{2} b_{V8} + b_{V10}\, , \qquad
\omega_2\equiv \frac{1}{2s^2_Z} b_{V10}
\end{align}
For simplicity, in (\ref{omdef1}) we have dropped the family indices, but one should keep in mind that the NLO corrections are in general different for electron and muon final states. Incidentally, notice that $\kappa_i$ also contain a universal (family-independent) contribution, proportional to $\beta_1$, $\delta_G$ and $a_{XU1}$, which results from taking into account the NLO renormalization effects. $\delta_G$ above stands for the renormalization of the Fermi constant, which includes 4-fermion operators not listed in (\ref{naive1}). More details can be found in \cite{Buchalla:2013wpa}.

For completeness we will also discuss the weakly-coupled case using the EFT 
developed in~\cite{Buchmuller:1985jz} using the notation of 
\cite{Grzadkowski:2010es} (for different approaches see~\cite{Passarino:2012cb,Pomarol:2013zra}). The relevant operators are now 
\begin{align}
{\cal{O}}_{HB}&= g^{\prime 2}B_{\mu\nu}B^{\mu\nu} H^{\dagger}H;&\qquad
{\cal{O}}_{HW}&=g^2\langle W_{\mu\nu}W^{\mu\nu}\rangle H^{\dagger}H\nonumber\\
{\cal{O}}_{HWB}&=gg^{\prime}H^{\dagger}W_{\mu\nu} H B^{\mu\nu};&\qquad
{\cal{O}}_{HD}&=|H^{\dagger}D_{\mu}H|^2\nonumber\\
{\cal{O}}_{Hl}^{(1)}&=(H^{\dagger}i\stackrel{\leftrightarrow}{D}_{\mu}H) ({\bar{l}}
\gamma^{\mu}l);&\qquad
{\cal{O}}_{Hl}^{(3)}&=(H^{\dagger}i\stackrel{\leftrightarrow}{D^a}_{\!\!\!\!\mu}H) 
({\bar{l}}\gamma^{\mu}\tau_a l)\nonumber\\
{\cal{O}}_{He}&=(H^{\dagger}i\stackrel{\leftrightarrow}{D}_{\mu}H) 
({\bar{e}}\gamma^{\mu}e);&\qquad
{\cal{O}}_{H\Box}&=(H^{\dagger}H)\Box(H^{\dagger}H)
\end{align} 
and (${\tilde{X}}_{\mu\nu}=\epsilon_{\mu\nu\lambda\rho}X^{\lambda\rho}$)
\begin{align}
{\cal{O}}_{H{\tilde{W}}}&=g^2\langle {\tilde{W}}_{\mu\nu}W^{\mu\nu}\rangle H^{\dagger}H;
\quad
{\cal{O}}_{H{\tilde{B}}}=g^{\prime 2}{\tilde{B}}_{\mu\nu}B^{\mu\nu} H^{\dagger}H;\quad
{\cal{O}}_{H{\tilde{W}}B}=gg^{\prime}H^{\dagger}{\tilde{W}}_{\mu\nu}H B^{\mu\nu}
\end{align}
for the CP-even and CP-odd sectors, respectively. 
The effect of ${\cal{O}}_{H\Box}$ is to renormalize the Higgs kinetic term. This 
shift can be absorbed by a field redefinition of $H$, which then affects 
the $H\to ZZ$ coupling. This is of no relevance for the distributions but 
affects the global normalization of the decay~\cite{Giudice:2007fh}. 
For comparison with the nonlinear case it is convenient to define 
${\bar{\alpha}}_j=v^2\alpha_j$. The result reads
\begin{align}\label{corrections2}
F_1&=\left(1+{\bar{\alpha}}_{H\Box}+\frac{{\bar{\alpha}}_{HD}}{4}-\delta_G\right)\nonumber\\
b_{2,3}&=\frac{e^2}{2}\left(2t_Z^2{\bar{\alpha}}_{HB,H{\tilde{B}}}+
t_Z^{-2}{\bar{\alpha}}_{HW,H{\tilde{W}}}+{\bar{\alpha}}_{HWB,H{\tilde{W}}B}\right)
\nonumber\\
b_{2,3}^{\gamma}&=e^2\left(-2t_Z {\bar{\alpha}}_{HB,H{\tilde{B}}}+
t_Z^{-1}{\bar{\alpha}}_{HW,H{\tilde{W}}}-
\frac{1}{2}(t_Z^{-1}-t_Z){\bar{\alpha}}_{HWB,H{\tilde{W}}B}\right)
\end{align}
and
\begin{align}\label{omdef2}
\kappa_1&\equiv 1+({\bar{\alpha}}_{Hl1}+
{\bar{\alpha}}_{Hl3}-{\bar{\alpha}}_{He})-
\frac{{\bar{\alpha}}_{HD}}{4}-\delta_G\nonumber\\
\kappa_2&\equiv 1-\frac{1}{2s^2_Z} {\bar{\alpha}}_{He}+
\frac{1}{c^2_Z-s^2_Z}\left(\frac{{\bar{\alpha}}_{HD}}{4}+
e^2\frac{{\bar{\alpha}}_{HWB}}{2s_Z^2}+\delta_G\right)\nonumber\\
\omega_1&\equiv 2({\bar{\alpha}}_{Hl1}+
{\bar{\alpha}}_{Hl3}-{\bar{\alpha}}_{He})\, ,\qquad
\omega_2\equiv -\frac{1}{s^2_Z} {\bar{\alpha}}_{He}
\end{align}
It is worth noting that, while the contributions to 
$h Z\ell^+\ell^-$ and $Z \ell^+\ell^-$, encoded in $\omega_i$ and $\kappa_i$, respectively,
come from the same (family-dependent) NLO operators, $\kappa_i$ also receives a universal NLO
renormalization through ${\cal{O}}_{HD}$, ${\cal{O}}_{HWB}$ and the operators associated with $\delta_G$. 
Therefore, the contact term contribution to $h\to Z\ell^+\ell^-$ is in general uncorrelated 
to $Z\to \ell^+\ell^-$, even in the case of the linearly-realized Higgs sector. 
Similarly, the $Z$ mass term and the $h\to ZZ$ vertex come from the same
LO operator but NLO corrections renormalize them differently. As a result, $\delta F_1\neq 0$ in (\ref{corrections2}).

\section{Observables and form factor determination}\label{sec:III}

In Section 2 we pointed out that at NLO there are 6 independent form factors 
entering the dynamical functions $J_i$. With high enough statistics one 
can fit the full distribution $J$ to experimental data. However, at least in 
the first stages of the run~2 at the LHC, where statistics will be rather 
limited, it is more efficient to devise a set of observables that can project 
out the different form factor combinations through angular asymmetries. 

A possible strategy is to extract $G_VG_A$ from the forward-backward asymmetry 
$A_{\alpha\beta}$ in $\alpha$ and $\beta$, after integration over $\phi$:  
\begin{align}\label{aab}
A_{\alpha\beta}
&=\left(\frac{d\Gamma}{ds}\right)^{-1} \int_{-1}^{1} d\!\cos\alpha\, {\mathrm{sgn}}(\cos\alpha)
 \int_{-1}^{1}d\!\cos\beta\, {\mathrm{sgn}}(\cos\beta)
\displaystyle\frac{d\Gamma}{ds\, d\!\cos\alpha \, d\!\cos\beta} =\frac{J_3}{J_1+J_2}
\end{align}
and $(G_V^2+G_A^2)$ from an asymmetry $A_\phi$ in the angle $\phi$:
\begin{align}\label{aph}
A_{\phi}
&=\left(\frac{d\Gamma}{ds}\right)^{-1}\int_0^{2\pi} d\phi\, {\mathrm{sgn}}(\cos 2\phi) 
\displaystyle\frac{d\Gamma}{dsd\phi} =\frac{32}{9\pi}\frac{J_9}{J_1+J_2}
\end{align}
Knowing $A_{\alpha\beta}$ and $A_{\phi}$, $H_{V,A}$ can be determined through the 
combinations $g_VH_V+g_AH_A\simeq g_AH_A$ and $g_VH_A+g_AH_V\simeq g_AH_V$. These can be extracted, 
respectively, from the total rate given in (\ref{dgamds}) and the 
asymmetry $B_\phi$,
\begin{align}\label{bph}
B_{\phi}
&=\left(\frac{d\Gamma}{ds}\right)^{-1}\int_0^{2\pi} d\phi\, {\mathrm{sgn}}(\cos \phi) 
\displaystyle\frac{d\Gamma}{dsd\phi} =\frac{\pi}{2}\frac{J_6}{J_1+J_2}
\end{align}
The observables discussed so far test new physics in the CP-even sector. 
CP-odd contributions are parametrized by $K_{V,A}$, which can be determined 
through the structures $g_VK_V+g_AK_A\simeq g_AK_A$ and $g_VK_A+g_AK_V\simeq g_AK_V$. 
They can be extracted from 2 additional asymmetries in $\phi$:
\begin{align}\label{cdph}
C_{\phi}
&=\left(\frac{d\Gamma}{ds}\right)^{-1}\int_0^{2\pi} d\phi\, {\mathrm{sgn}}(\sin 2\phi) 
\displaystyle\frac{d\Gamma}{dsd\phi} =\frac{32}{9\pi}\frac{J_8}{J_1+J_2} \nonumber\\
D_{\phi}
&=\left(\frac{d\Gamma}{ds}\right)^{-1}\int_0^{2\pi} d\phi\, {\mathrm{sgn}}(\sin\phi) 
\displaystyle\frac{d\Gamma}{dsd\phi} =\frac{\pi}{2}\frac{J_4}{J_1+J_2}
\end{align}
Similar CP-odd observables have been discussed previously in the literature~\citer{Soni:1993jc,Sun:2013yra}.

In order to assess the experimental relevance of these asymmetries, 
we will rely on numerical estimates of new-physics effects based on general 
power-counting arguments. Accordingly, one would naively expect the NLO 
coefficients given in the previous section to be generically of 
${\cal{O}}(v^2/\Lambda^2)$, 
with $\Lambda\sim 4\pi v$. Therefore, keeping track of the gauge couplings, 
we will assume $F_1=a+{\cal{O}}(v^2/\Lambda^2)$, 
$g_{V,A}=g_{V,A}^{(0)}+g{\cal{O}}(v^2/\Lambda^2)$, 
$b_{2,3}^{(\gamma)}\sim e^2{\cal{O}}(v^2/\Lambda^2)$ and
$h_{V,A}\sim g {\cal{O}}(v^2/\Lambda^2)$. 

The main source of deviations from the 
SM comes from $a$ in $F_1$. This parameter measures the signal strength 
of $h\to ZZ^*$, and is currently constrained to deviate less than $20\%$ from 
the SM. Since our conclusions will be independent of it, we will set $a=1$ and $F_1=1$ for 
simplicity. New-physics corrections are then naturally dominated by 
$\delta g_{V,A}$ and $h_{V,A}$. $\delta g_{V,A}$ are constrained by the $Z$ partial width and LEP data
sets bounds on them at the $10^{-3}$ level~\cite{Han:2004az,Pomarol:2013zra}, which is within the 
EFT expectation. $h_{V,A}$ are instead unconstrained, and might in principle
attain values larger than the naive EFT dimensional 
estimate because of numerical enhancements. Consider, for instance, the local $h\to Z\ell^+\ell^-$ couplings
$h_{V,A}$ to be induced by the tree-level exchange of a composite heavy 
vector resonance $R$, mediating $h\to Z R^*$, $R^*\to\ell^+\ell^-$. 
Then $h_{V,A}\sim v^2/M^2_R\sim v^2/\Lambda^2$. If
$M_R$ is numerically smaller than $\Lambda\approx 3\,{\rm TeV}$ by a factor of
three, say, the resulting value of $h_{V,A}$ might be 5-10 times bigger 
than the naive EFT estimate. This assumes consistency with other
phenomenological constraints, which is plausible in view of the 
free parameters in this scenario.

\begin{figure}[t]
\begin{center}
\includegraphics[width=6.5cm]{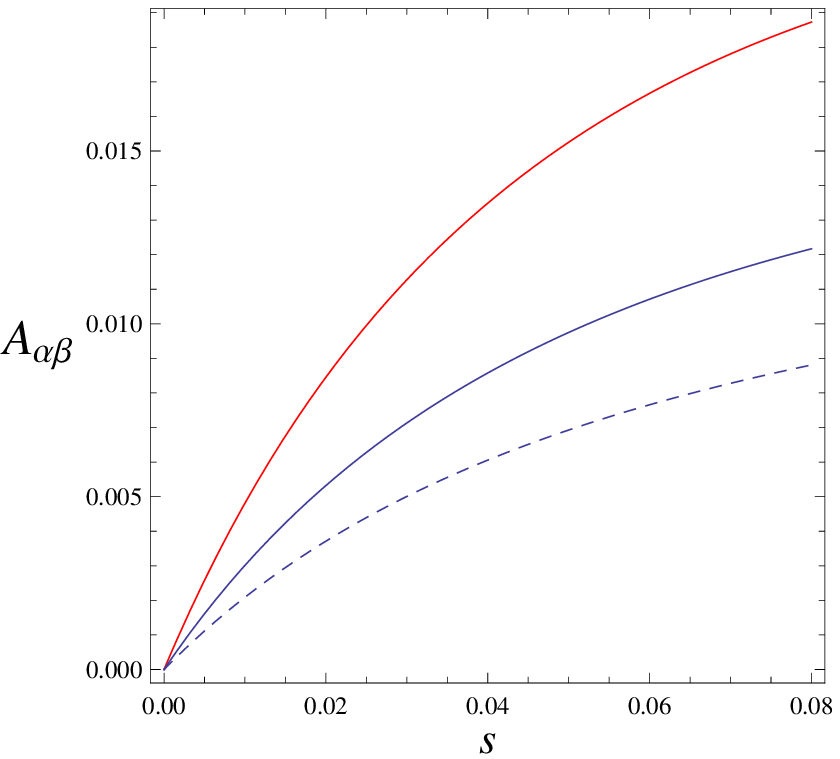}
\hskip 1.0cm
\includegraphics[width=6.5cm]{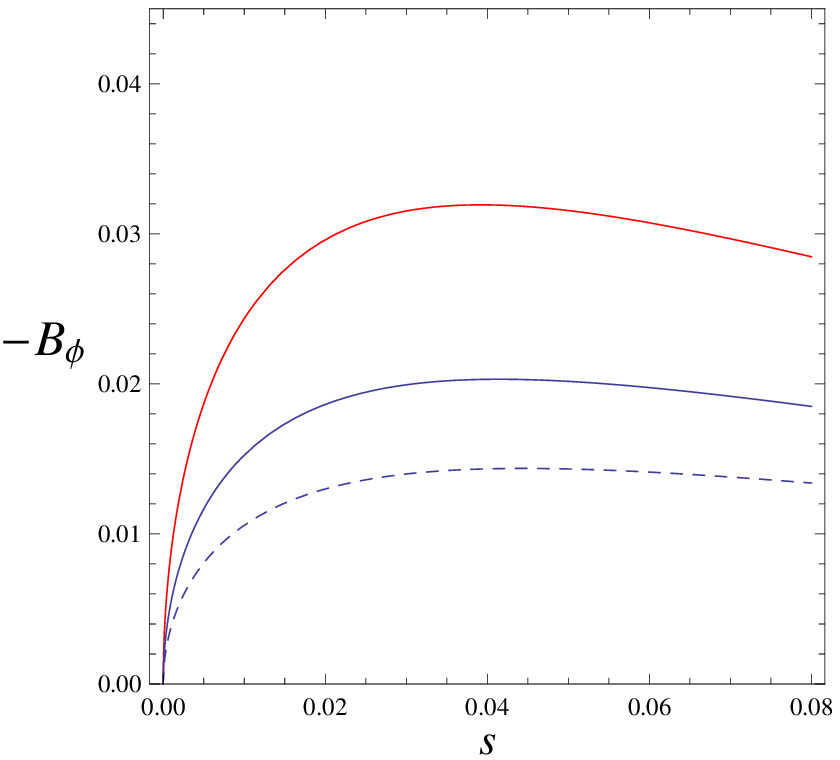}
\vskip 0.5cm
\includegraphics[width=6.5cm]{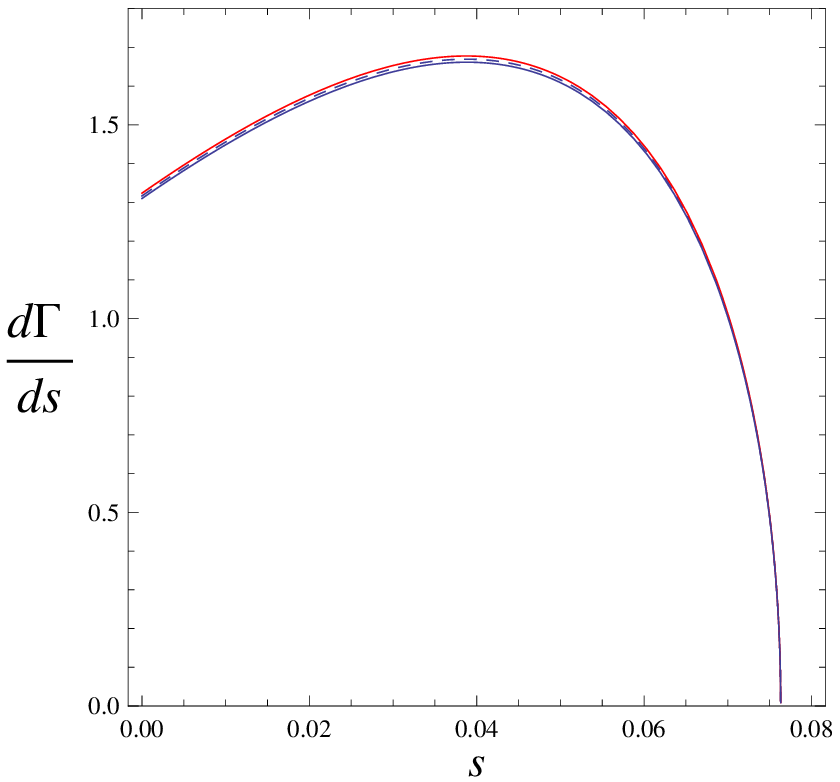}
\end{center}
\caption{\small{\it{Values for the angular asymmetries $A_{\alpha\beta}$ and 
$B_{\phi}$ defined in the main text. The dashed line corresponds to the 
SM prediction, while the solid lines incorporate potential new-physics 
effects for the parameter choices  
$(h_V,h_A)=v^2/\Lambda^2(-2,0.3)$ (in blue) and 
$(h_V,h_A)=v^2/\Lambda^2(-6,0.3)$ (in red). 
For comparison, the lower panel shows the differential mass distribution 
(in units of $10^{-6}$ GeV). 
The plots illustrate the high sensitivity of the angular asymmetries to 
new physics for scenarios where the mass distribution is left almost 
unaffected.}}}\label{fig:numerics}
\end{figure}

For simplicity we will consider a scenario where $h_{V,A}\neq 0$, with all other corrections set to zero. 
Due to the smallness of $g_V$ in the SM, the most sensitive probes of new physics are those linear in 
$G_V$, namely $A_{\alpha\beta}$ and $B_{\phi}$, with corrections that 
can easily reach 50-100\%. Incidentally, notice that neither 
$A_{\alpha\beta}$ nor $B_{\phi}$ are constrained by the angular distributions 
collected for the spin-parity analysis~\cite{Aad:2013xqa}. This has to be compared with the mass distribution, 
with typical corrections of a few $\%$. However, both corrections are 
uncorrelated. Qualitatively, $h_V$ controls $A_{\alpha\beta}$ and $B_{\phi}$ while $h_A$ affects the mass 
distribution. Thus, one can get large corrections on the former while barely 
affecting the latter. In Fig.~\ref{fig:numerics} we illustrate 
such scenarios for the parameter choices 
$(h_V,h_A)=v^2/\Lambda^2(-2,0.3)$ and 
$(h_V,h_A)=v^2/\Lambda^2(-6,0.3)$. 

With the LHC running at 14 TeV and with an integrated luminosity of 3000 fb$^{-1}$, 
one expects around 6400 reconstructed events for $h\to Z\ell^+\ell^-$~\cite{Anderson:2013fba}. 
With such statistics one could in principle reach a $1-2\%$ sensitivity in the observables
that we are discussing. Since the overall effects for $A_{\alpha\beta}$ and $B_{\phi}$ lie around 
the $\%$ level, as illustrated in Fig.~\ref{fig:numerics}, they could be accessible
at the LHC, at least in its final stage. Regarding the CP-odd sector, within the range of validity 
of our EFT, the asymmetries $C_{\phi}$ and $D_{\phi}$ are expected to be below the 
per-mille level and thus clearly out of reach for detection at the LHC. 

These estimates could be made more precise by analysing the size of the backgrounds 
associated to the specific angular dependences. Such an analysis goes 
beyond the scope of the present paper, but naively they should be 
substantially reduced as compared to the total decay 
rate~\cite{Bolognesi:2012mm,Gainer:2011xz,Chen:2012jy}. 
In this case, $A_{\phi}$ might turn out to be especially suited to extract
$(G_V^2+G_A^2)$ with higher precision than through the total decay rate. 

Before closing this section, one should note that, strictly speaking, the form factors $G_i,H_i,K_i$ always appear in 
combination with $g_{V,A}$ in the products
\begin{align}
&(G_V^2+G_A^2)(g_V^2+g_A^2);&\qquad &(G_VG_A)(g_Vg_A)\nonumber\\
&(G_VH_V+G_AH_A)(g_V^2+g_A^2);&\qquad &(G_VH_A+G_AH_V)(g_Vg_A)\nonumber\\
&(G_VK_V+G_AK_A)(g_V^2+g_A^2);&\qquad &(G_VK_A+G_AK_V)(g_Vg_A)
\end{align}
which account for the processes $h\to Z\ell^+\ell^-$ and $Z\to \ell^{\prime +}\ell^{\prime -}$, respectively.
In order to determine $G_i,H_i,K_i$ with a certain precision, $g_{V,A}$ should be known comparably well.
Unfortunately, with LEP data the bounds on $g_V$ and $g_A$ are too loose to be 
informative~\cite{Han:2004az}. In contrast, the ILC could offer a clean determination
of the $Ze^+e^-$ couplings, since the center-of-mass enhanced corrections to $W^+W^-$ production can be cast 
entirely in terms of these corrections~\cite{Buchalla:2013wpa}. As a result, they get 
singled out at high energies and, within the ILC energy-range, they can naturally be boosted to a $20\%$ 
correction to the production cross-section. An analogous mechanism for $Z\mu^+\mu^-$ couplings could in principle be 
pursued in a muon linear collider through $\mu^+\mu^-\to W^+W^-$.


\section{Conclusions}\label{sec:V}

We have studied, in a general and systematic way, how the decay 
$h\to Z\ell^+\ell^-$ can be used to probe for physics beyond the
Standard Model in the Higgs sector. For this purpose we have employed 
a general parametrization of the amplitude in terms of form factors,
neglecting lepton masses. In view of the large gap between the electroweak scale 
and the expected scale of new physics, an effective field theory approach appears to be 
the most efficient tool. We have computed the form factors in terms of the
coefficients of an effective Lagrangian, which is defined by the SM gauge 
symmetries, a light scalar singlet $h$ and the remaining SM particles, but      
is otherwise completely general. 

The main points of our analysis can be summarized as follows.
\begin{itemize}
\item
We discuss the most general observables arising from the full angular
distribution of the 4-lepton final state in $h\to Z\ell^+\ell^-$, 
$Z\to\ell^{'+}\ell^{'-}$. The 9 coefficients $J_i$ describing the
angular distribution are expressed through the 6 form factors
$G_{V,A}$, $H_{V,A}$ and $K_{V,A}$.
 
\item
Interesting observables, besides the dilepton-mass spectrum $d\Gamma/ds$, 
can be constructed from the angular distribution. Examples are:
\begin{itemize}
\item The forward-backward asymmetry $A_{\alpha\beta}$ measuring $J_3$ and $B_{\phi}$ measuring $J_6$.
These quan\-ti\-ties are strong\-ly sup\-pressed in the SM because of the
smallness of the vectorial coupling  $g_V$. On the other hand, this implies
an enhanced relative sensitivity to new physics. The required precision  
of a few $\%$ might be within reach of the LHC. 
\item $J_7$ or $J_9$ give similar information as $d\Gamma/ds$,
but should have different experimental systematics because of the
characteristic angular dependence associated with them.
\item
CP violation in the coupling of $h$ to electroweak bosons
is probed by $J_4$, $J_5$, $J_8$, which enter the terms in the decay 
distribution odd in the angle between the dilepton planes $\phi$. Their effects
are however expected at the per-mille level and thus out of reach of the LHC. 
\end{itemize}
\item
The form factors are expressed in terms of the coefficients
of the complete effective Lagrangian at next-to-leading order,
${\cal{O}}( v^2/\Lambda^2\sim 1/(16\pi^2))$. We use the electroweak
chiral Lagrangian, extended to include a light Higgs singlet $h$,
and take into account all NLO new-physics effects at tree level,
including the renormalization of SM fields and parameters.
The effective Lagrangian for a linearly realized Higgs is also
considered with operators up to dimension 6.  
\item
Based on effective-theory power counting, the potentially dominant 
impact of new-physics arises from the 
leading-order $hZZ$ coupling $a$, which only affects the overall
decay rate, but not the angular and dilepton-mass distributions.
The latter can only be modified by the NLO coefficients in the Lagrangian. 
\item
Power counting gives a typical size of the NLO coefficients of
$\sim v^2/\Lambda^2\sim 1\%$, up to coupling
constants and numerical factors. With this estimate the new-physics effects
are typically small. In particular, the contributions of the virtual 
$Z$ and $\gamma$, which could in principle be inferred from the profiles of 
the different mass distributions turn out to be at the 
per-mille level and therefore too small to be detected. Somewhat larger effects 
(up to 5\%) may be possible in specific scenarios, for instance from 
enhanced $hZ\bar ll$ local couplings $h_{V,A}$ in a strongly-interacting Higgs sector.
Quantities such as $A_{\alpha\beta}$ and $B_{\phi}$, with their large sensitivity to
NP corrections, could be especially interesting in this respect. 
\item
For the quantitative extraction of new-physics coefficients from data,
radiative corrections have to be taken into account.
To NLO (one loop) in the Standard Model they have been computed in
\cite{Bredenstein:2006rh,Kniehl:2012rz}.
\end{itemize}

New-physics effects in $h\to Z\ell^+\ell^-$ decay distributions are expected to be small, even in the 
case of a strongly-interacting Higgs sector. The tree level SM contribution is the 
dominating effect and NP can potentially show up typically at the percent 
level. Nevertheless, this NP suppression can be compensated by statistics, 
and we have shown that interesting opportunities exist 
for precision measurements, already at the LHC, 
which could provide valuable insight into electroweak symmetry breaking. 
The rich subject of $h\to Z\ell^+\ell^-$ observables should therefore
be fully explored by experiment.
 
\section*{Acknowledgments}
We thank Elvira Rossi, Mario Antonelli and Wolfgang Hollik for useful discussions.
O.~C.~wants to thank the University of Naples for very pleasant stays 
during the different stages of this work. O.~C.~is supported in part by 
the DFG cluster of excellence 'Origin and Structure of the Universe' and the ERC Advanced Grant project 
'FLAVOUR' (267104).
G.~D'A.~is grateful to the Dipartimento di Fisica of Federico II University, 
Naples, for hospitality and support and acknowledges partial support by 
MIUR under project 2010YJ2NYW.
\\

\appendix

\section{4-body decay kinematics}\label{sec:app}

\begin{figure}[t]
\begin{center}
\includegraphics[width=10.0cm]{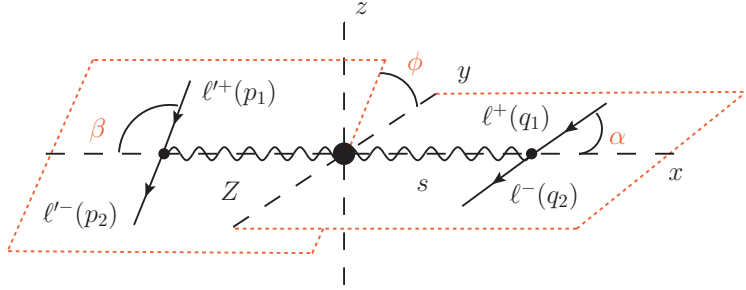}
\end{center}
\caption{\small{\it{Definition of angles in 
$h\to Z\ell^+\ell^-$, $Z\to\ell^{'+}\ell^{'-}$.}}}\label{fig:angledef}
\end{figure}

In order to describe the full angular distribution of 
$h(k)\to Z(p)\ell^+(q_1)\ell^-(q_2)$, followed by 
$Z(p)\to\ell^{'+}(p_1)\ell^{'-}(p_2)$, one needs to specify 4 variables. 
A convenient choice is to select the invariant mass of the dilepton pair,
$q^2=(q_1+q_2)^2\equiv M^2_h s$, together with 3 angles $\alpha,\beta,\phi$. 
The angular variables are defined as in \cite{Cabibbo:1965zz,Pais:1968zz}: $\alpha$ and 
$\beta$ are the angles, in the respective dilepton c.m.s., between the 
${\ell}^+$ momenta and the direction of the dilepton systems as seen from 
the Higgs rest frame, while $\phi$ is the angle between the dilepton planes. 
Refering to the $xyz$-coordinate frame shown in Fig. \ref{fig:angledef}, 
the precise definition of the angles can be stated as follows:
\begin{itemize}
\item
The dilepton momentum $\vec q$ in the rest frame of $h$ defines
the direction of the positive $x$-axis, $\vec e_x=\vec q/|\vec q|$. 
\item
$\alpha$ is the angle between $\vec e_x$ and the $\ell^+$ momentum
$\vec q_1$ in the $\ell^+\ell^-$ c.m.s.
\item
$\beta$ is the angle between $-\vec e_x$ and the $\ell^{'+}$ momentum
$\vec p_1$ in the $\ell^{'+}\ell^{'-}$ c.m.s.
\item
$\phi$ is the relative angle between the normals of the decay planes,
$\vec e_x\times \vec q_1/|\vec e_x\times \vec q_1|$ and
$\vec e_x\times \vec p_1/|\vec e_x\times \vec p_1|$, counted positive from the
former to the latter in the positive direction around $\vec e_x$. 
\end{itemize}

In the following we will assume that the final-state leptons are massless, 
which is a very good approximation at the electroweak scale.
The lepton momenta in the respective dilepton centre-of-mass systems can then be 
pa\-ra\-me\-trized as
\begin{align}
q^\mu_{1}&=\frac{M_h\sqrt{s}}{2}\left(1,{\hat{n}}_1\right),\qquad 
q^\mu_{2}=\frac{M_h\sqrt{s}}{2}\left(1,-{\hat{n}}_1\right)\nonumber\\
p^\mu_{1}&=\frac{M_h\sqrt{r}}{2}\left(1,{\hat{n}}_2\right),\qquad 
p^\mu_{2}=\frac{M_h\sqrt{r}}{2}\left(1,-{\hat{n}}_2\right)
\end{align}
with the unit vectors
\begin{equation}
{\hat{n}}_1=(\cos\alpha,\sin\alpha,0)\, ,\qquad
{\hat{n}}_2=(-\cos\beta,\sin\beta\cos\phi,\sin\beta\sin\phi)
\end{equation}
The range of the kinematical variables is
\begin{align}\label{kinevar}
0  \, &\le \, s\, \le \,(1-\sqrt{r})^2 = 0.076\nonumber\\
0\,&\le \, \alpha,~\beta \, \le \, \pi\nonumber\\
0 \, &\le \, \phi \, \le \, 2\pi
\end{align}
The momenta can be boosted to the Higgs rest frame with the following 
velocities:
\begin{equation}
\beta_q=\frac{\lambda}{1-r+s}\, ,\qquad
\beta_p=\frac{\lambda}{1+r-s}
\end{equation}
where $r$, $s$ and $\lambda$ are defined in (\ref{rslamdef}). The relevant kinematical invariants are then given by  
\begin{align}
q_1\cdot p_1&=\frac{M^2_h}{8}\Big[\left(1+c_{\alpha}c_{\beta}\right)g_H+
\lambda(c_{\alpha}+c_{\beta})-2\sqrt{r s}\, s_{\alpha}s_{\beta}c_{\phi}\Big]
\nonumber\\
q_1\cdot p_2&=\frac{M^2_h}{8}\Big[\left(1-c_{\alpha}c_{\beta}\right)g_H+
\lambda(c_{\alpha}-c_{\beta})+2\sqrt{r s}\, s_{\alpha}s_{\beta}c_{\phi}\Big]
\nonumber\\
q_2\cdot p_1&=\frac{M^2_h}{8}\Big[\left(1-c_{\alpha}c_{\beta}\right)g_H-
\lambda(c_{\alpha}-c_{\beta})+2\sqrt{r s}\, s_{\alpha}s_{\beta}c_{\phi}\Big]
\nonumber\\
q_2\cdot p_2&=\frac{M^2_h}{8}\Big[\left(1+c_{\alpha}c_{\beta}\right)g_H-
\lambda(c_{\alpha}+c_{\beta})-2\sqrt{r s}\, s_{\alpha}s_{\beta}c_{\phi}\Big]\nonumber\\
q_1\cdot q_2&=\frac{M^2_h}{2} s\, ,\qquad
p_1\cdot p_2=\frac{M^2_h}{2} r\, ,\qquad 
\epsilon_{\mu\nu\lambda\rho}p_1^{\mu}p_2^{\nu}q_1^{\lambda}q_2^{\rho}=
\frac{M^4_h}{8}\lambda \sqrt{r s}\, s_{\alpha}s_{\beta}s_{\phi}
\end{align}
where $g_H\equiv 1-r-s$, $c_\chi\equiv\cos\chi$, $s_\chi\equiv\sin\chi$ and 
$\epsilon_{0123}=+1$.


\end{document}